\documentstyle[aps,preprint,prl]{revtex}

\newcommand {\rv}{\vec{r}} 
\newcommand {\rhv}{\vec{\rho}}

\begin{document}
\draft

\title{Binding energy of localized biexcitons in quantum wells.  }

\author{C. Riva,\cite{clara} F. M. Peeters,\cite{francois} and
  K. Varga.\cite{varga} \ \
\\Departement Natuurkunde, Universitaire Instelling Antwerpen,
Universiteitsplein 1, B-2610 Antwerpen, Belgium  } \date{\today}
\maketitle
\begin{abstract}

A variational calculation of the ground state energy of a biexciton 
in a GaAs/AlGaAs quantum well is presented. The well width
fluctuations leading to trapping of the biexcitons are modeled by
a parabolic potential. The results obtained for different well widths
 are compared with recent experimental data. 
Good agreement is obtained
both for the biexciton binding energy and for the Haynes factor.
We find that the structure of a biexciton is similar to the one of the
$H_2$
molecule.

\end{abstract}


 
In the last few years considerable experimental progress was made in
detecting biexcitons in semiconductor systems. A biexciton is a system
consisting
of two excitons which are bound together. Since the first observation of
biexcitons in quantum wells reported by
R.C. Miller {\em et al.},\cite{Miller} there have been many studies,
both
experimental\cite{Miller,first-experimental,Pantke,Bar,Kim,Smith,Birkedal} 
and theoretical,\cite{Kleinman,different-ass,Bastard} on this subject.
Kleinman\cite{Kleinman} developed a variational model that gives
results in agreement with the first experimental
\cite{Miller,first-experimental} data. However later
experimental studies,
carried out with more advanced techniques, have reported substantial larger
values for  the binding
energy\cite{Pantke,Bar,Kim,Smith,Birkedal} as compared to the early
experiments. 
For example the  Haynes factor, $f_H = E_b^{XX}/E_b^X$, which is the
ratio between the biexciton binding energy ( $ E_b^{XX} $) and the
exciton binding energy ($ E_b^X $),   found by
Birkedal {\em et al.}\cite{Birkedal} has a value in the range 0.19-0.22 for well widths
between 80 and 160$\dot{A}$, while Kleinman predicts a value  in the
range 0.11-0.12. In order to explain  this difference between theory
and experiment, calculations were
carried out with different techniques as well
as with new assumptions\cite{different-ass} 
on the spatial form of the biexciton. Singh {\em et
  al.}\cite{different-ass}
assumed a square-like arrangement of the electrons and the holes in a
two dimensional 
biexciton which resulted in
$f_H=0.228$.  
The latter  approach is rather {\em ad hoc} and does not include the
finite thickness of the biexciton and consequently is not able to
explain the well width dependence of the biexciton binding energy.

The aim of the present paper is to explain  the recent experimental results 
by considering localization effects on the biexciton.
This localization can be  a consequence 
of  the modulation of the thickness of the quantum well.
 Indeed  for a quantum well
of width $L$ a variation in well width  of $\Delta L$  produces a change
in the zero point energy of the order  of $ \Delta L
\hbar^2\pi^2/(mL^3)$.
For  a quantum well of 160\AA {} a fluctuation of about  2.5\AA
{} induces a
zero point fluctuation of the order of 0.5 meV which compares to a
biexciton binding energy of  about 1.5meV.


Using the effective mass approximation a biexciton in a quantum well
can  be described by the Hamiltonian

\begin{equation}
\hat{H}_{XX}= \hat{H}_{1X}+\hat{H}_{2X}+\sum_{i=e,h}(-1+2(
\delta_{i,e}+\delta_{i,h} ) ) {e^2 \over  |\rv_{1i}-\rv_{2i}|}+
V_{conf}(z)+\sum_ {i=e,h}\sum_{j=1,2}{1\over 2}m_i \omega^2 \rho_{j,i}^2,
\label{firstham}
\end{equation}
with
\begin{equation}
\hat{H}_{iX}= -{\hbar^2 \over 2 m^*_e}\nabla^2_{ie}-{\hbar^2 \over 2
  m^*_h}\nabla^2_{ih} - {e^2 \over  |\rv_{ie}-\rv_{ih}|},
\end{equation}
where the indexes 1, 2  indicate the first and second
exciton, $m^*_i$ is the effective mass of the particle {\em i}, and
$V_{conf}(z)$ is the confining potential associated with the presence of
the quantum well. $\omega$ is the frequency of the shallow parabolic 
 confining potential in the quantum well plane that models the quantum well
width fluctuations and $\rhv$ is the projection of $\rv$ in the plane orthogonal to the well axis.
The confinement energy 
is much larger  than the biexciton and
exciton binding energy which allows us
to treat the system as a
quasi-two dimensional system, i.e. we can separate the contribution to the wave
function along the quantum well axis, chosen as z-axis, from the
contribution along the plane, the $\rhv$-plane,
\begin{equation}
\Psi(\rv_{1e},\rv_{2e},\rv_{1h},\rv_{2h})={\mathcal F }
(z_1,z_2,z_a,z_b)\psi(\rhv_{1e},\rhv_{2e},\rhv_{1h},\rhv_{2h}). 
\end{equation}
The component of the wave function along the z-axis is taken as
a product of the 1D ground state wave functions for an electron (hole)
in a hard wall quantum well.
Averaging the Hamiltonian over the z-component ${\mathcal
  F}(z_{1e},z_{2e},z_{1h},z_{2h})=f_e(z_{1e})f_e(z_{2e})f_h 
 (z_{1h})f_h(z_{2h})$  we obtain the following 
effective 2D Hamiltonian 
\begin{equation}
\begin{array}{rl}
\hat{H}_{\rho}=&\displaystyle\frac{1}{1+\sigma}(\sigma \Delta_a+\sigma
\Delta_b+\Delta_1 +\Delta_2 )-2(U_{1,a}+U_{1,b}+U_{2,a}+U_{2,b}-U_{a,b}-U_{1,2}) 
 \\ \\
  & \displaystyle +{1\over 4} 
(1+\sigma) [ {1 \over \sigma} \omega^2
(\rho_a^2+\rho_b^2+\rho_1^2+\rho_2^2)],
\end{array}
 \label{secondham}
\end{equation}
where $\sigma=m_e/m_h$ is the mass ratio between the electron and the
hole and $U_{i,j}$ 
is the effective Coulomb potential obtained by averaging the real Coulomb
potential over the wave functions along the z-direction. 
In Eq.(\ref{secondham}) we expressed  the
energy in units of   
$R_y=e^2/2\epsilon a_B$ and the length in units of 
$a_B=\epsilon \hbar^2/e^2 \mu$
with $\epsilon$ the static dielectric constant and $\mu$ the in-plane reduced mass
of the electron-hole system. Using $\sigma=0.68$, i.e. $m_e/m_0=0.067$,
$m_h/m_0=0.099$, $\epsilon=12.1$ for a GaAs/AlGaAs with concentration
of Al=25\%, we find $R_y=3.7meV$ and
$a_B=160\dot{A}$.

It has been shown\cite{Platzman} that $U_{i,j}$ can be well approximated by
$1/ \sqrt{\lambda^2+\rho^2}$, where $\lambda=0.2 L$ with $L$ the width of
the well which is valid for hard well confinement. The latter approach
is a very good approximation for the wide quantum wells considered in
the present paper. Using this approximation
the Hamiltonian (\ref{secondham}) was solved with the stochastical
variational technique of Ref.12 with the trial wave
function taken as a 
combination of correlated Gaussian functions,
\begin{eqnarray}
&\psi =\sum_{n=1}^K \Phi_{nLs},&
\\
&\Phi_{nLs}={\mathcal A}\{ \chi_{SM_s}& Y_{LM_L}(\sum_{i=1}^{3}u_{ni}
    \vec{\zeta}_i) exp(-{1 
    \over 2} \sum_{i,j=1}^3 A_{nij} \vec{\zeta}_{i} \cdot
  \vec{\zeta}_{j}) \},
\label{gaussian}
\end{eqnarray}
where $ \vec{\zeta}_1$ and $\vec{\zeta}_2$ are the distance vectors
between the hole and the electron in the first and in the second
exciton respectively and
$\vec{\zeta}_3$ is the distance between the centers of mass of the two
excitons, $\chi_{SM_S}$ is the spin function, $Y_{LM_L}$ is the
angular function and 
${\mathcal A}$ is the antisymmetrization operator.
The interaction among the different particles is taken into account via
the non diagonal terms of the matrix {\bf A}. The best set of
variational parameters $\{u_{ni}, A_{nij}\} $ is found using a
stochastical method. The dimension of the basis $K$ is 
increased until the required
accuracy is achieved. Here we are only interested in  the ground state, and
consequently, the total 
angular momentum L and the total spin S are zero.

The biexciton binding energy is obtained as follows
\begin{equation}
E^{XX}_b=2E^{X}-\tilde{E}^{XX}
\label{binding-formula}
\end{equation}
where $\tilde{E^{XX}}=(E^{XX}-4*\omega)$ 
is the biexciton ground state energy  
as referred to  the
 four free particle in the shallow parabolic confinement
potential
and  $E^{X}$ the
ground state of  a {\em mobile} exciton. 

The quantum well width fluctuations ($\Delta L$) shift the zero point
energy  of the electrons and holes by  $ \Delta L
\hbar^2\pi^2/(mL^3)$ and  thus it is reasonable to assume that  the
shallow confinement 
$\omega$ is inversely dependent on some power of the well width 
$L$. These considerations suggest to search for  such
a dependency in the form of $\omega(L)= a/L^n$. 
In order to do this we considered
the experimental data reported in various experiments
\cite{Pantke,Bar,Kim,Smith,Birkedal} and we used $\omega$ as a fitting
parameter. 
 The obtained confinement frequencies are
plotted in Fig. \ref{omega.vs.L}. 
The influence of the confinement on the biexciton binding energy is
shown in the inset of Fig. \ref{omega.vs.L} for different values of
the quantum well width. Note that $E_b^{XX}$ increases almost linear
with $\omega$.
On the basis of the above zero point energy 
argument we expect a $\omega \propto L^{-3}$
which seems to agree with the experimental results for $L/a_B
<0.7$. Noting that there is a lot of scatter between the different
experimental results it seems that  the best overall behaviour of
$\omega$ is given by  $\omega
=0.06/ L^{-1} $, although a constant value of 
 $\omega= 0.068 R_y \approx 0.26 meV $  also agrees with the
 experimental results, at least for $L/a_B >0.6$

In Fig. \ref{expdata-res} we plot 
our $\omega=0$ biexciton binding energy are comparable to
those found by Kleinman although our biexciton and exciton  energy
are considerably smaller. Adding a shallow confinement potential in
the quantum well plane increases the biexciton binding energy
substantially (about a factor of 2) and brings the theoretical results
 in agreement with the
experimental results. We show our results for a constant confinement
of $\hbar \omega=0.068 R_y$ and for a well width dependent confinement
of $\hbar \omega/R_y=0.06/(L/a_B)$. The latter gives a better overall agreement
with the experimental data.
 The different experimental results are from different quantum wells
 which have been not grown under the same conditions  and consequently
 the well width fluctuations can also be substantially different.

 The Haynes factor, which is the ratio between the biexciton energy and
 the exciton energy, is found experimentally to be almost independent
 on the width of the well. Our theoretical results,
 see Fig. \ref{Haynes}, seem to confirm this and leads to 
 $f_H \approx .22$ which compares to the value, $f_H=0.228$, found by
 Singh {\it  et al.}
Although our theoretical results show a weak well width dependence
they fall withing the scatter of the experimental results.
 weak well width dependence but  

Note that the previous calculation by Kleinman results into $f_H=0.13$
which is about a factor of two smaller than the one found
experimentally and very close to the value we found in the case $\omega=0$.
 
In order to investigate the structure of the biexciton we evaluate the
average distance between the two electrons, between the two holes
and between the electron and hole which is defined as follows
\begin{equation}
<\rho_{ij}>=\int |\psi(\rhv_{1e},\rhv_{2e},\rhv_{1h},\rhv_{2h})|^2 |\rhv_i-\rhv_j|
d\rhv_{1e} d\rhv_{2e} d\rhv_{1h} d\rhv_{2h},   
\end{equation}
with i,j =1e,2e,1h,2h. The results are depicted  in the inset of 
 Fig. \ref{average-distances} as function of the well width.
Note that the average electron-electron and the average hole-hole 
distances
 are comparable, and the average electron-hole distance is
such that $<\rho_{eh}>/<\rho_{ee}>\approx 1.35$
For a square 2D biexciton as assumed by Singh {\em at al.} one has
$\rho_{ee}=\rho_{hh}=\sqrt{2}\rho_{eh}$. Noticing that this equation
is satisfied within  4\% one may naively believe that the
electrons and the holes are situated on the corner of a square.

Next  we consider   the pair
correlation function for the electron-hole pair
\begin{equation}
P_{eh}(\rho)={1 \over
  2}\sum_{i=1e,2e}\sum_{j=1h,2h}<\delta(|\rhv_i-\rhv_j|-\rho)>,   
\end{equation}
and the one of the electron-electron ( hole-hole) pair
\begin{equation}
P_{^{ee}_{hh}}(\rho)={1\over
  2}\sum_{i=^{1e,2e}_{1h,2h}}\sum_{j=^{1e,2e}_{1h,2h}} 
<\delta(|\rhv_i-\rhv_j|-\rho)> , 
\end{equation}
which is plotted in  Fig. \ref{average-distances}. Note  that the
electron is much strongly 
correlated to the hole and that there is a high probability for the two
particles to stay very close to each other. While  electrons (holes) stay
quite far from each other. This result argues against the model of a 
square biexciton proposed by Singh {\em et al.} and suggests that the electrons
and holes orbit around each other like in single excitons and that the centers
of mass of the two
excitons are  a certain distance apart which is approximately equal to the
average  hole-hole (electron-electron) distance. Such a
configuration is similar to the one of a $H_2$ molecule. 


In conclusion, we found that in order to explain the experimentally
avalaible results on the biexciton binding energy in quantum wells we
have to assume that the biexcitons are trapped. The trapping potential
is assumed to be parabolic which models the trapping potential induced
by the well width fluctuations found in  real systems. Our results
indicate that the trapping potential frequency  has a smaller well
width dependence  than expected  from a pure monolayer well width
behaviour, except maybe for the quantum wells which are  smaller than 100 \AA. The
Haynes factor is practically independent from the well width in
agreement with the experimental results. By investigating the
interparticle correlation functions we found that the biexciton can be
considered like a $H_2$ molecule rather than a square arrangement of
electrons and holes as proposed by Singh {\em et al.}.

\section{Acknowledgment}

Part of this work is supported by  
the Flemish Science
Foundation (FWO-Vl) and the `Interuniversity Poles of Attraction
Program - Belgian State, Prime Minister's Office - Federal Office for
Scientific, Technical and Cultural Affairs'. F.M.P. is a Research
Director with FWO-Vl.

\begin{figure}
\caption{The confinement trapping frequency as function of
 of the quantum well width L. 
  The different symbols are the results obtained from the fitting to
  the experimental biexciton binding energy.
The different curves shows the inverse power laws with a=0.06 , b=0.03
and c=0.02 and L measured in Bohr radii. 
The inset shows the dependence of the biexciton  binding energy on the
confinement fo different well width.}
\label{omega.vs.L}
\end{figure}


\begin{figure}
\caption{Comparison between different  theoretical results for the binding
  energy of the biexciton (curves) and experimental data (symbols).}
\label{expdata-res}
\end{figure}

\begin{figure}
\caption{ The Haynes factor is plotted versus the well width.
 The dashed curve represents the results from the theory of
  Kleinman. The different symbols are the experimental results from
  different groups. }
\label{Haynes}
\end{figure}


\begin{figure}
\caption{The different pair correlation functions, for a
biexciton in quantum well of width $ L/a_B= 1 $ and confinement energy
$\hbar \omega/R_y = 0.068$, with $\rho_{ij}=|\rhv_i-\rhv_j|$. 
 The inset shows the average distance between the different particles
 in the biexciton as function of the well width for a confinement
 energy $\hbar \omega/ R_y= 0.068$.}
\label{average-distances}
\end{figure}





\begin{references}

\bibitem [*] {clara} Electronic address: riva@uia.ua.ac.be.
\bibitem [\dag] {francois} Electronic address: peeters@uia.ua.ac.be.
\bibitem [\ddag] {varga} Permanent address: Institute of Nuclear Research
of the Hungarian Academy of Science (ATOMKI), Debrecen, H-4001, Hungary.
\bibitem{Miller} R.C. Miller, D.A. Kleinman, A.C. Gossard, and
  O. Munteanu, Phys. Rev. B {\bf 25}, 6545 (1982).
\bibitem{Kleinman} D. Kleinman, Phys. Rev. B {\bf 28}, 871 (1983).
\bibitem{first-experimental} S. Charbonneau, T. Steiner,
  M.L.W. Thewalt, E.S. Koteles, J.Y. Chi, and B. Elman, Phys. Rev. B
  {\bf 38}, 3583 (1988); R. Cingolani, Y. Chen, and K. Ploog,
  Phys. Rev. B {\bf 38}, 13478 (1988); D.C. Reynolds, K.K. Bajaj,
  C.E. Stutz, R.L. Jones, W.H. Theis, P.W. Yu, and K.R. Evans,
  Phys. Rev. B {\bf 40}, 3340 (1989). 
\bibitem{Pantke} K.H. Panthke, D. Oberhauser,
  V.G. Lyssenko, J.M. Hvam, and G. Weimann, Phys. Rev. B {\bf 47},
  2413 (1993).
\bibitem{Bar} S. Bar-Ad and I. Bar-Joseph, Phys. Rev. Lett. {\bf 68},
  349 (1992).
\bibitem{Kim} J. Kim, D. Wake, and J. Wolfe, Phys. Rev. B {\bf 50},
  15099 (1994).
\bibitem{Smith} G. Smith, E. Mayer, J. Kuhl, and K. Ploog, Solid State
  Commun. {\bf 92}, 325 (1994).
\bibitem{Birkedal} D. Birkedal, J. Singh, V.G. Lyssenko, J. Erland, and
  J.M. Hvam, Phys. Rev. Lett. {\bf 76}, 672 (1996).
\bibitem{different-ass} J. Singh, D. Birkedal, V.G. Lyssenko, and
  J.M. Havm, Phys. Rev. B {\bf 53}, 15909 (1996).
\bibitem{Bastard} O. Heller,Ph. Lelong,and G. Bastard, Phys. Rev. B {\bf
    56}, 4702 (1997).
\bibitem{Platzman} R. Price, X. Zhu, S. Das Sarma, and P.M. Platzman,
  Phys. Rev. B {\bf 51}, 2071 (1995)
\bibitem{Varga1} K. Varga and Y. Suzuki, Phys. Rev. C {\bf 52}, 2885
  (1995); Phys. Rev. A {\bf 53}, 1907 (1996).
\end{references}
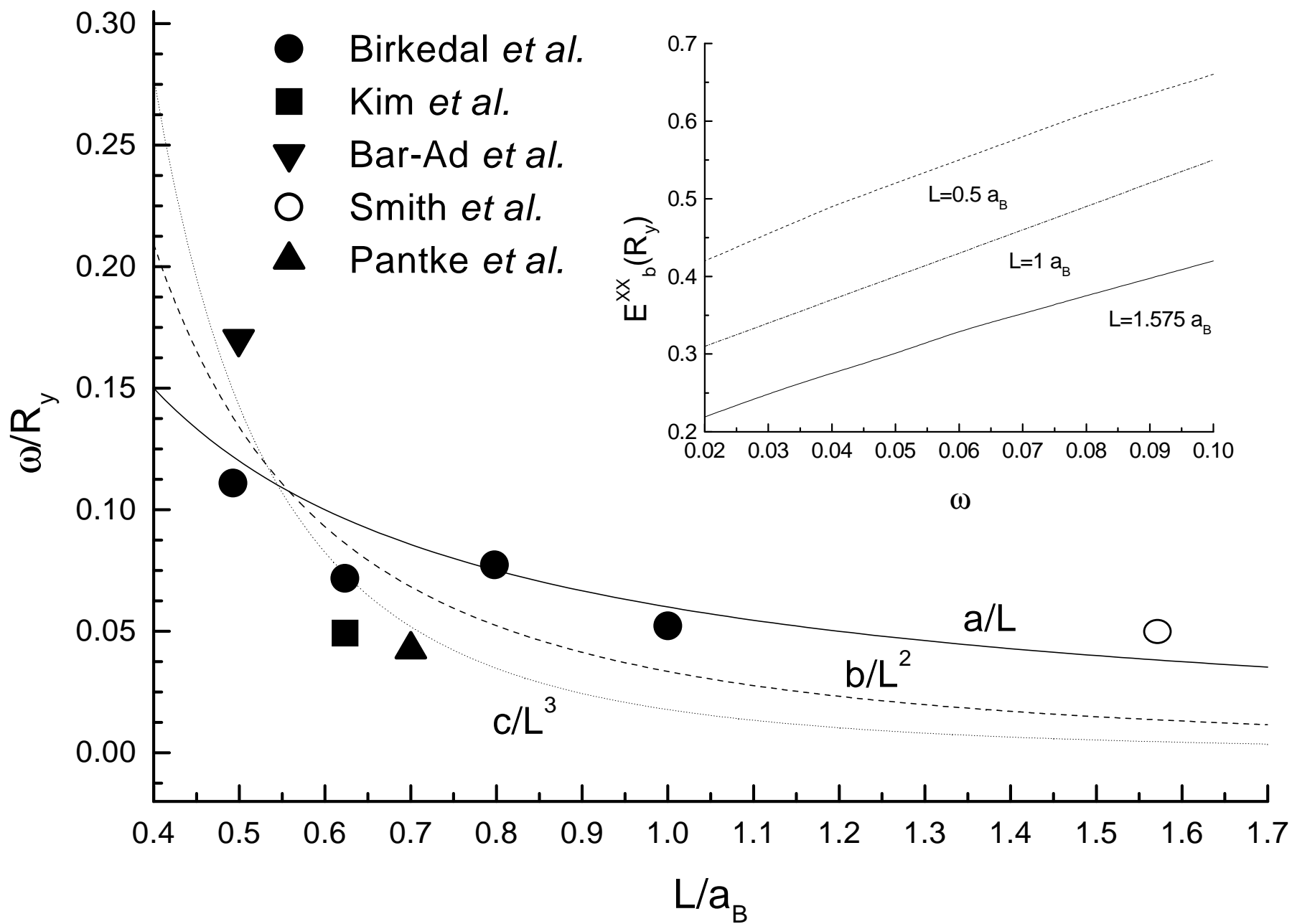
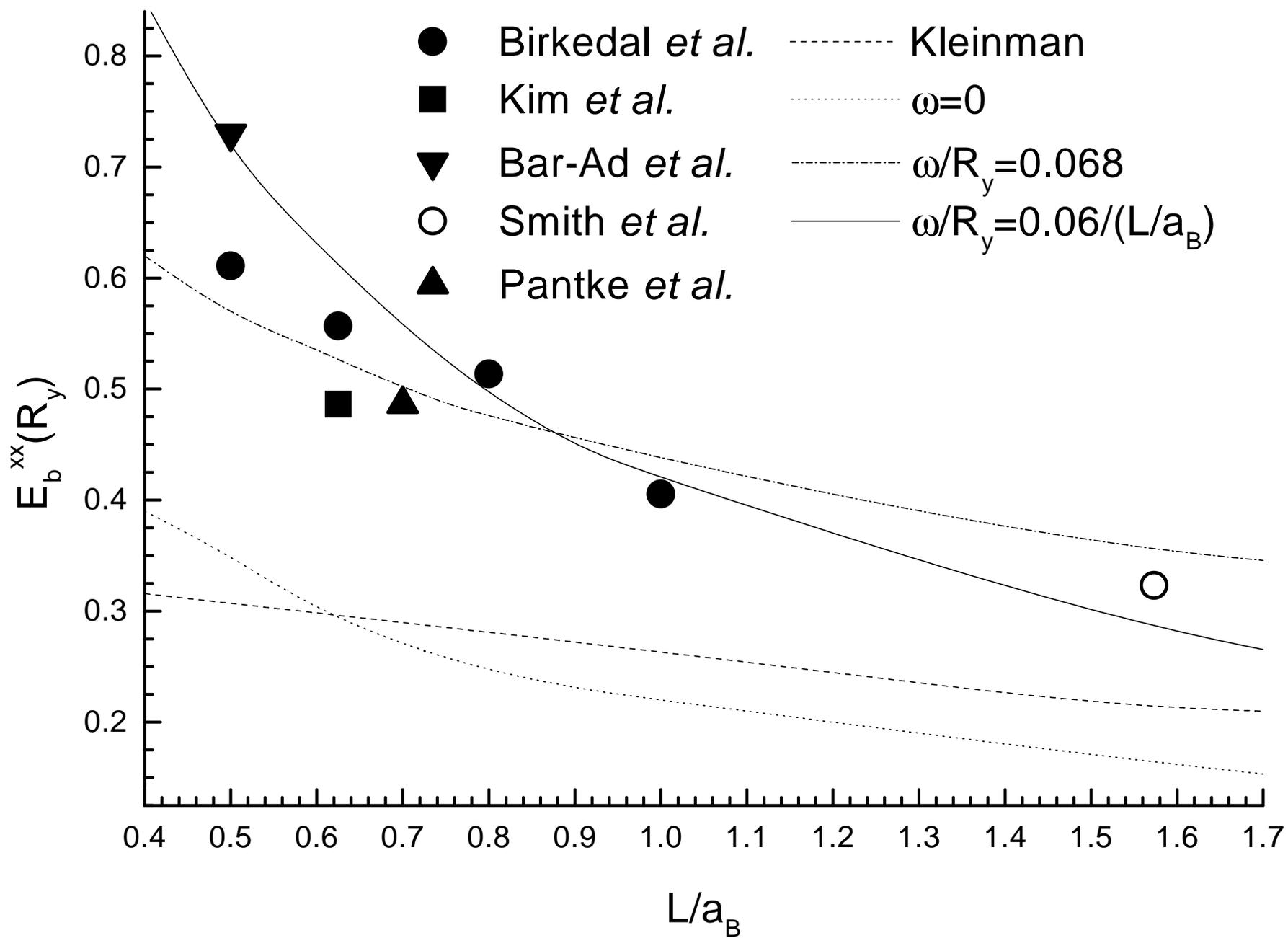
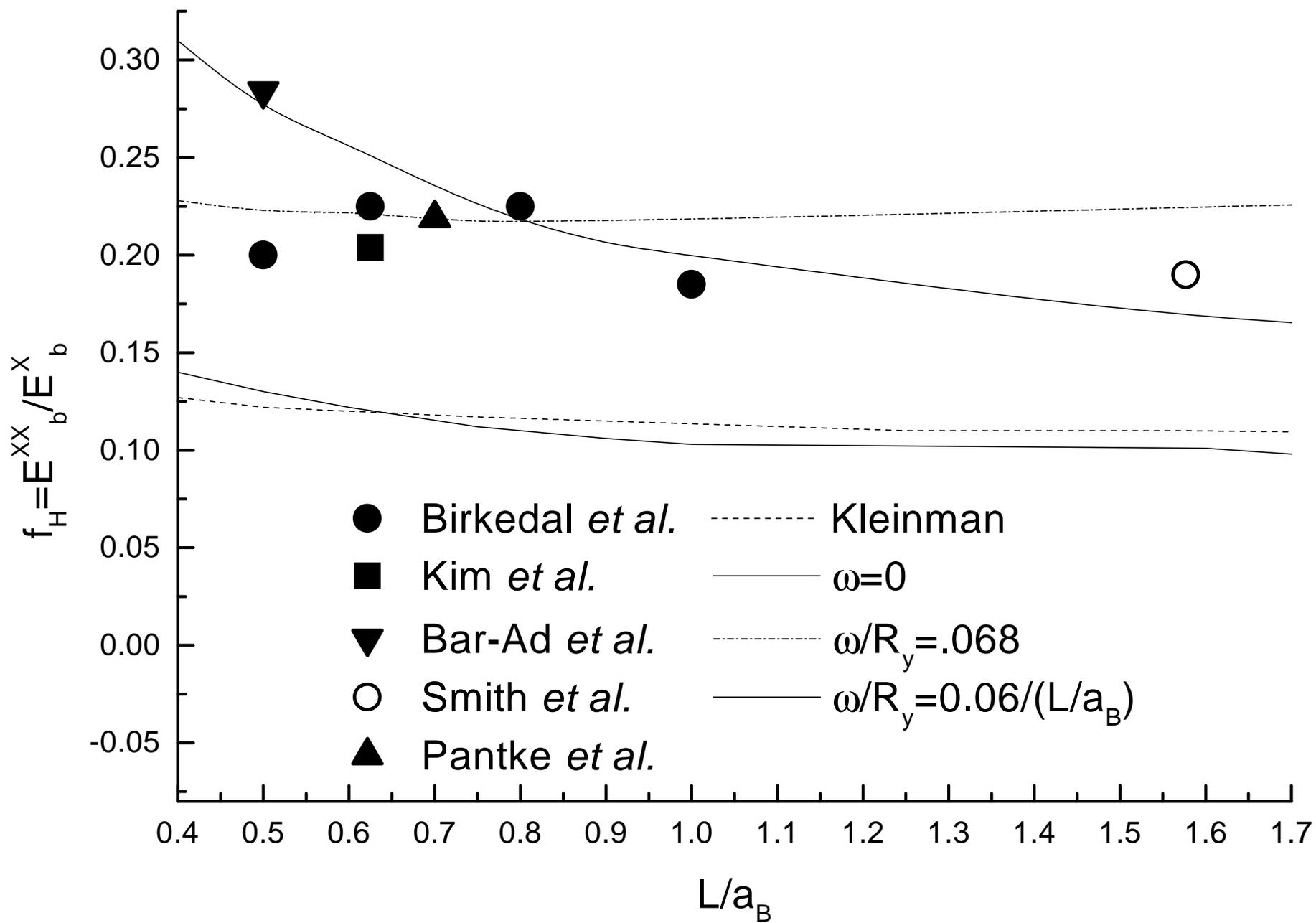
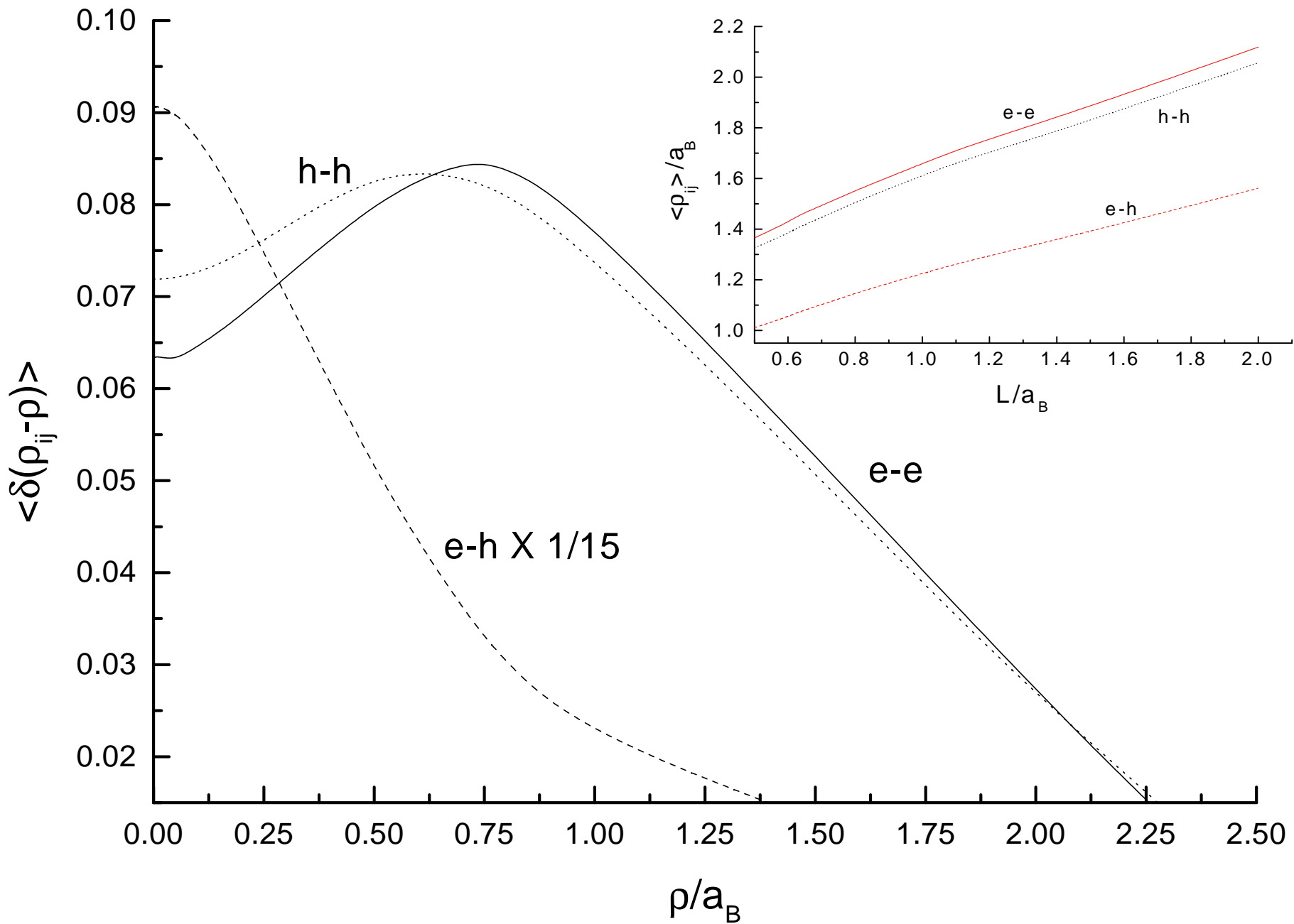
\end{document}